\documentclass[]{article}
\usepackage{multicol}
\usepackage{apj}
\newcommand{\mincir}{\raise -2.truept\hbox{\rlap{\hbox{$\sim$}}\raise5.truept
\hbox{$<$}\ }}
\newcommand{\magcir}{\raise -2.truept\hbox{\rlap{\hbox{$\sim$}}\raise5.truept
\hbox{$>$}\ }}
\newcommand{\siml}{\raise -2.truept\hbox{\rlap{\hbox{$\sim$}}\raise5.truept
\hbox{$<$}\ }}
\newcommand{\simg}{\raise -2.truept\hbox{\rlap{\hbox{$\sim$}}\raise5.truept
\hbox{$>$}\ }}

\newcommand{\be}{\begin{equation}}
\newcommand{\ee}{\end{equation}}
\newcommand{\ba}{\begin{eqnarray}}
\newcommand{\ea}{\end{eqnarray}}
\newcommand {\h} {$h^{-1} \, Mpc \,$}

\newcommand {\msun} {$h^{-1} \  M_{\odot} \;$}

\newcommand{\vel}{\,{\rm km\,s^{-1}}}

\begin{document}


\vspace{15mm}
\begin{center}
\uppercase{The Observational Mass Function of Nearby Galaxy Clusters}\\
\vspace*{1.5ex}
{\sc Marisa Girardi$^{1,2}$, Stefano Borgani$^{3}$,
Giuliano Giuricin$^{2}$,
Fabio Mardirossian$^{1,2}$,\\
and Marino Mezzetti$^{2}$}\\
\vspace*{1.ex}
{\small
$^1$Osservatorio Astronomico di Trieste, Via Tiepolo 11, I-34131 Trieste, Italy\\
$^2$SISSA, via Beirut 4, I-34014 Trieste, Italy; and
Dipartimento di Astronomia, Universit\`{a} degli Studi di Trieste, Trieste, Italy\\  
E-mail: girardi, giuricin, mardiros, mezzetti @sissa.it \\
$^3$INFN, Sezione di Perugia, c/o Dipartimento di Fisica
dell'Universit\`a,\\ via A. Pascoli, I-06123 Perugia, Italy;
E-mail: borgani@perugia.infn.it}
\end{center}
\vspace*{-6pt}

\begin{abstract}
  We present a new determination of the mass function of galaxy
  clusters, based on optical virial mass estimates for a large sample of 152
  nearby ($z\le 0.15$) Abell--ACO clusters, as provided by Girardi et
  al. (1998). This sample includes both data from the literature and
  the new ENACS data. The resulting mass function is reliably
  estimated for masses larger than $M_{lim}\simeq 4\times
  10^{14}$\msun, while it is affected by sample incompleteness at
  smaller masses. 

  We find $N(>M_{lim})=(6.3\pm 1.2)\,10^{-6}$ $(h^{-1}
  Mpc)^{-3}$ for cluster masses estimated within a 1.5 \h~radius.  Our
  mass function is intermediate between the two previous estimates by
  Bahcall \& Cen (1993) and by Biviano et al. (1993).

  Based on the Press--Schechter approach, we use this mass function to
  constrain the amplitude of the fluctuation power spectrum at the
  cluster scale.  After suitably convolving the PS predictions with
  observational errors on cluster masses and {\sl COBE}--normalizing
  the fluctuation power spectrum, we find $\sigma_8=(0.60\pm 0.04)
  \Omega_0^{-0.46+0.09\Omega_0}$ for flat low--density models and
  $\sigma_8=(0.60\pm 0.04) \Omega_0^{-0.48+0.17\Omega_0}$ for open
  models (at the 90\% c.l.).


%
\vspace*{6pt}
\noindent   
{\em Subject headings: }
galaxies: clusters: general -
cosmology: observations - cosmology: theory - large scale
structure of universe.

\end{abstract}

\begin{multicols}{2}

\section{INTRODUCTION}

According to the scenario of hierarchical formation of cosmic
structures, clusters of galaxies arise from rare high peaks of the
initial density fluctuation field and, at present time, represent the
largest virialized cosmic structures. As a consequence, standard analytical
arguments based on the Press \& Schechter (1974; hereafter PS)
approach show that their abundance is highly sensitive to the
amplitude of the density fluctuations on the cluster mass scale (e.g.,
White, Efstathiou \& Frenk 1993). In particular, if $M$ is the typical
cluster mass and $\Omega_0$ is the cosmic density parameter, then the
typical size of the density fluctuation collapsing into a cluster is
$R\propto (\Omega_0 M)^{1/3}$. Typical cluster masses, $M\sim 5\times
10^{14}$\msun \footnote{Here $h$ is the Hubble constant in units of
100$\vel$Mpc$^{-1}$}, are rather close to the average mass contained
within a sphere of 8\h~ radius. Therefore, it is common to
express constraints from the observed local cluster mass function
(i.e., the number density of clusters of a given mass), $n(M)$, in
terms of $\Omega_0$ and $\sigma_8$, the
r.m.s. density fluctuation within a sphere of 8\h~ radius (e.g., 
Eke, Cole, \& Frenk 1996; Viana \& Liddle 1996).

  From the theoretical side, the mass function of clusters is easily
obtainable either from N--body simulations or, to a comparable degree
of reliability (e.g., Lacey \& Cole 1993; Eke et al. 1996; Borgani et
al. 1998), by applying the analytical PS recipe. On the contrary,
predictions of cosmological models about quantities which are more
directly connected to observations, like the internal velocity
dispersion of member galaxies, $\sigma_v$, and the $X$--ray
temperature of the intracluster gas, $T_X$, are much less
straightforward and rely on assumptions which are quite debated. As
for $\sigma_v$, N--body simulations allow reliable predictions of the
internal velocity dispersion of clusters, but only for the DM
particles (e.g., Frenk et al. 1990; Borgani et al. 1997; van Haarlem,
Frenk, \& White 1997), while the connection to the observed $\sigma_v$
requires a plausible model of galaxy formation.
Numerical simulations by Frenk et al. (1996) have shown that present
uncertainties about the physics of galaxy formation in clusters can
lead to an incorrect determination of the virial mass by up to a factor
five.  As for $T_X$, it is generally believed to be a rather accurate
indicator of the cluster mass (e.g., Evrard, Metzler, \& Navarro 1996).
On the other hand, both analytical predictions and hydrodynamical
cluster simulations, based only on the adiabatic physics of the
intracluster gas, are able to reproduce neither the observed relation
between bolometric $X$--ray luminosity and temperature (e.g., 
Eke, Navarro \& Frenk 1997; Bryan \& Norman 1998,
 and references therein), nor
the observed size of core radii in the gas distribution (e.g., Makino,
Sasaki, \& Suto 1998, and references therein).  Such a discrepancy
creates a need to introduce additional physics to explain the
thermodynamics of the the intra--cluster gas (e.g., Bower 1997;
Cavaliere, Menci, \& Tozzi 1997, and references therein).

   From the observational side, cluster masses are inferred from
either $X$--ray or optical data, under the general hypothesis of
dynamical equilibrium.  Estimates based on gravitational lensing do
not require assumptions about the dynamical status of the cluster, but
a good knowledge of the geometry of the potential well is necessary
(e.g., Narayan \& Bartelmann 1997).  Claims for a discrepancy (by a
factor 2-3) between cluster masses obtained with different methods
casted doubts about the general reliability of mass estimates (e.g.,
Wu \& Fang 1997).  However, recent analyses have shown that such
discrepancies can be explained by the different way in which strong
cluster substructures bias mass estimates based on different methods
(Allen 1997; Girardi et al.  1997). In particular, the analysis by
Girardi et al. (1998, hereafter G98) demonstrated that, for nearby
clusters without strong substructures, a good overall agreement exists
between $X$--ray and optical virial mass estimates.

A reliable determination of cluster masses is a necessary but not
sufficient condition to guarantee a reliable estimate of the mass
function. Indeed, the latter also requires a cluster sample
which is large enough to be a fair representation of the cluster
population in the local Universe, at least above a certain mass limit.
In past years, such requirements limited the possibility of obtaining a
stable result about the cluster mass function and, in fact, the only
two estimates so far presented (Bahcall \& Cen 1993, hereafter BC93;
Biviano et al. 1993, hereafter B93) turn out to be quite discrepant.
BC93 derived cluster masses by using global scaling relations to
connect mass to cluster richness and $X$--ray temperature. B93
estimated cluster masses from a homogeneous virial analysis of about
70 clusters and found a larger abundance, by about one
order of magnitude, at the mass scale $M\simeq 5\times 10^{14}$\msun
(cf. also Figure~3 below).

A new and more robust determination of the observational mass function
based on optical virial analysis is now possible thanks to the recent
availability of a large amount of redshift data for rich clusters, as
provided by the ESO Nearby Abell Cluster Survey (ENACS; Katgert et al.
1996; 1998).  Recently, G98 computed virial masses in a homogeneous
way for a large sample of $\sim 170$ clusters, obtained by combining
the ENACS sample with data from other authors (see also Fadda et al.
1996; hereafter F96). The extension of this sample makes it the
largest available data set over which to reliably estimate the cluster
mass function.

In the following analysis we will use the G98 results with the twofold
aim of {\em (a)} estimating the optical virial mass function of nearby
galaxy clusters, and {\em (b)} investigating the subsequent
constraints on the power spectrum of cosmic density fluctuations.

This paper is organized as follows. In \S~2 we briefly describe the
statistical sample of clusters masses as obtained by G98 and then
present the determination of the cluster mass function. Afterwards, we
compare it with previous determinations based on optical data by BC93
and B93. In \S~3 we apply the PS approach to place constraints on the
fluctuation power spectrum for different cosmological models. We give
a brief summary of our main results and draw our conclusions in \S~4.

\section{THE DETERMINATION OF THE MASS FUNCTION} 
\subsection{The Statistical Sample of Cluster Masses}

In the following analysis we largely use the results by G98, 
who analyzed optical data (galaxy positions and
velocities) for 170 nearby clusters having $z\le0.15$. The original
optical data partly comes from the literature and partly from the
ENACS dataset (Katgert et al. 1996; 1998). In this section we
summarize the G98 results which are relevant for the mass function
analysis and we refer to that paper for further details.

The total mass of each cluster is computed by applying the virial
theorem to the member galaxies, under the usual assumption that the
mass distribution follows the galaxy distribution (e.g., Limber \&
Mathews 1960; Giuricin, Mardirossian, \& Mezzetti 1982; Merritt
1988; B93).  This assumption is supported by several pieces of
observational evidence, coming both from optical (e.g., Carlberg, Yee,
\& Ellingson 1997) and X-ray data (e.g., Watt et al. 1992; Durret et
al. 1994; Cirimele, Nesci, \& Trevese 1997), as well as from
gravitational lensing data (e.g., Narayan \& Bartelmann 1997).
Although some level of uncertainty still remains about the very
central cluster regions, detailed optical analyses (Cirimele et al.
1997; see also Merritt \& Tremblay 1994 for the Coma cluster) suggest
a very peaked galaxy distribution, in agreement with data on the dark
matter distribution from gravitational lensing (see, e.g., Narayan \&
Bartelmann 1997).  

In order to fit the galaxy distribution for each
cluster, G98 used a King--like profile, $\rho(r)\propto
(1+r^2/r_c^2)^{-3\beta/2}$, where $r_c$ is the core radius and $\beta$
is a free exponent. They found that, on average, $\rho(r) \propto
r^{-2.4}$ for the scaling of the galaxy number density in the outer
part of the clusters, thus in agreement with previous studies (e.g.,
Bahcall \& Lubin 1994; Girardi et al. 1995), with a peaked
distribution in the inner part ($r_c\sim 50\,h^{-1}$kpc; see also \S~8 of
G98).

G98 computed cluster masses within the radius $0.002\cdot \sigma_v$
\h, that represents a rough estimate of the virialization radius
$R_{vir}$, within which dynamical equilibrium is expected to hold.  A
surface term correction has also been applied to the usual virial mass
estimator which, otherwise, would have overestimated the true cluster
mass (e.g., The \& White 1976; Binney \& Tremaine 1987; Carlberg et
al. 1997b). This correction
amounts to $19\%$, very similar to that suggested by Carlberg et al.
(1997b) for CNOC clusters.

G98 compared their optical mass estimates to those derived from
$X$--ray analyses for a list of 66 clusters, which they compiled from
the literature, and found an overall good agreement (cf. Figure~5 of
G98).  This agreement is expected in the framework of the two
assumptions that mass follows the galaxy distribution and that
clusters are not far from dynamical equilibrium (see also 
Evrard et al. 1996; Schindler 1996; Carlberg et al. 1997c).
Only about $10\%$ of clusters show at least
two strongly superimposed peaks in their velocity distribution thus
indicating that they probably are far from dynamical equilibrium.  The
ambiguity in the treatment of these clusters -- peaks can be treated
as either disjoined or joined into a single structure -- leads to mass
estimates which differ by a factor $\sim 3$ on average (cf. \S~6 of
G98). The effect of these clusters with uncertain dynamics in the
estimate of the mass function will be discussed in the following.

As a concluding remark, we point out that the size of our sample,
along with the overall good agreement between $X$--ray and
optical cluster masses, makes it the largest available data set
over which to reliably estimate the cluster mass function.

\subsection{From Masses to the Mass Function}

The results from G98 that we will use in our analysis of the cluster
mass function are the following: {\em (i)} the values of the
(line-of-sight) velocity dispersions $\sigma_v$ ($\sigma_P$ in G98);
{\em (ii)} the values of the ``corrected'' virial masses, $M$
($M_{CV}$ in G98), contained within the radius $0.002\cdot \sigma_v$
\h; {\em (iii)} the King--like galaxy number--density profile which is
fitted for each cluster and used to rescale $M$ at different radii.
Hereafter, we limit our analysis to the 152 Abell--ACO clusters which
span the wide interval of richness classes $-1\le R \le 4$ (cf. Tab.~1
of G98), where the richness class $R=-1$ is assigned to clusters with
Abell number counts $N_C<30$ (Abell, Corwin, \& Olowin 1989).

The G98 sample is complete neither in volume nor in richness.
However, there exists a well defined, although rather scattered,
correlation between cluster richness and mass. Indeed, in Table~1 we
show that, although the cluster mass increases with richness, there is
a large mass superposition between different richness classes $R$ at a
given mass.  This is also explicitly shown in Figure~1, where we plot
the mass distribution for different $R$ classes.  Mazure et al. (1996,
\S~6; M96 hereafter) showed that the remarkable scatter in the
$\sigma_v$--$R$ relation is largely intrinsic. The propagation of this
scatter is the main reason for the scatter in the $M$--$R$ relation
since the relation between mass and $\sigma_v$ is rather strict (with
a scatter of $\sim$10\%; see the upper panel of Figure~2).  Therefore,
we expect that the scatter in the $M$--$R$ relation is also largely
intrinsic, much like that in the $\sigma_v$--$R$ relation.

\vspace{6mm} 
\hspace{-4mm}
\begin{minipage}{9cm}
\renewcommand{\arraystretch}{1.2}
\renewcommand{\tabcolsep}{1.2mm}
\begin{center}  
\vspace{-3mm}
TABLE 1\\
\vspace{2mm}
{\sc $M$ vs. Abell Richness $R$\\}
\footnotesize
\vspace{2mm}

%
%

\begin{tabular}{rrr}
\hline \hline
\multicolumn{1}{c}{$R$}
&\multicolumn{1}{c}{Cluster $\#$}
&\multicolumn{1}{c}{$log(M/M_{\odot})$}
\\
\hline
-1 & 7 & $14.08\pm0.15$\\
0  & 20& $14.13\pm0.10$\\
1  & 74&$14.24\pm0.06$\\
 2 & 35&$14.61\pm0.08$\\
$\ge$3&7&$14.74\pm0.11$\\
\hline
\end{tabular}


\newpage

\end{center}
{\footnotesize\parindent=3mm
The 17 multipeaked clusters (see text) are not reported
  here.}

\vspace{3mm}
\end{minipage}

In order to obtain a more representative cluster sample which also
accounts for the presence of this scatter in the $M$--$R$ relation, we
resample our clusters so as to mimic the $R$--distribution of the
Edinburgh--Durham Cluster Catalogue (EDCC), which is claimed to be
complete also for rather poor clusters (Lumsden et al. 1992).  We
compute the mass distribution by using 10,000 random extraction of the
measured $M$ values, distributed according to the EDCC
$R$--distribution, by rescaling the EDCC richness to Abell's 
(see also B93; Girardi et al. 1993; F96).  The
effectiveness of this resampling procedure has been shown by F96, who
applied it to a sample similar to the G98 one in order to obtain the
$\sigma_v$--distribution. As a result (cf. their Figure~6), they were able
to reproduce the $\sigma_v$--distribution derived by M96 for the volume-- and
richness--complete ENACS sample, with the further advantage of
extending the richness range from $R=1$ to $R=-1$ and, therefore, the
$\sigma_v$ range of completeness from $\sigma_v\ge 800 \vel$ down to
$\sigma_v \ge 650 \vel$.

\includegraphics{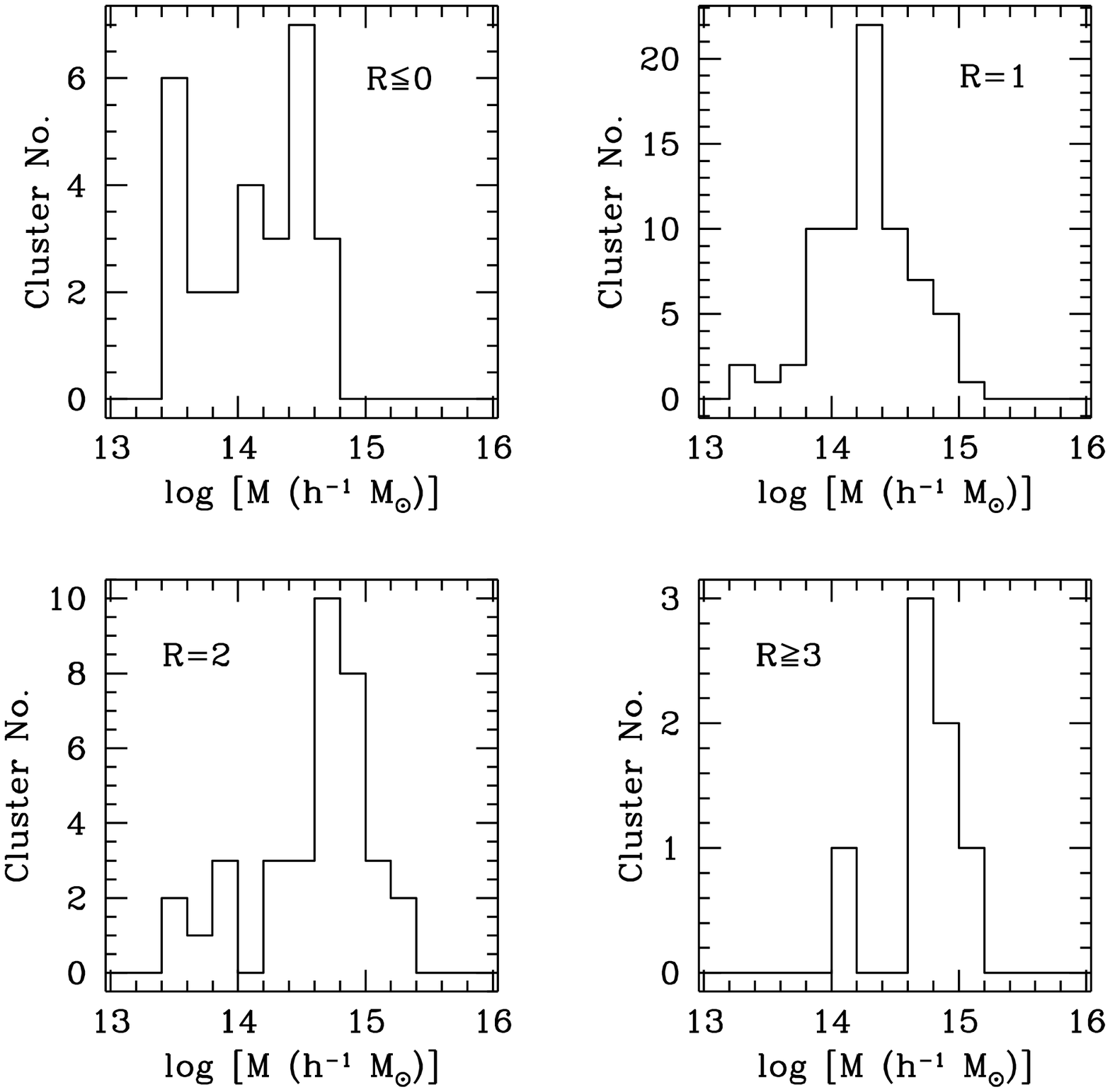}
$\ \ \ \ \ \ $\\
\vspace{9.4truecm}
$\ \ \ $\\
{\small\parindent=3.5mm {Fig.}~1.---
The mass distributions for clusters belonging to
  different richness classes. $R=0$ and $R=-1$ classes are here
  considered together, owing to their similar mass distribution and to
  the small number of $R=-1$ clusters (see also text and Table 1).
}
\vspace{5mm}

In the present analysis, we apply the resampling procedure both to the
whole sample of 152 $R\ge-1$ clusters and to the sample of 120 $R\ge1$
clusters and obtain the corresponding mass distributions.  Since no
substantial difference exists between masses for $R=-1$ and $R=0$
classes and owing to the small number of $R=-1$ clusters, we treat
these two classes together. We verified that no difference is found in
the final results if the two classes are instead treated separately.

Having determined the shape of the mass function through this
resampling procedure, we have still to fix its amplitude by resorting
to an external normalization.  To this purpose, we adopt the cluster
volume density $\bar N=8.6\times 10^{-6}$(\h)$^{-3}$ for $R\ge1$
clusters, as provided by M96 for ENACS, scaled to the EDCC
$R$-frequencies.  Since the M96 value was corrected for the
incompleteness of the Abell--ACO catalog with respect to the EDCC
catalog, our normalization is consistent with the resampling
procedure.

\includegraphics{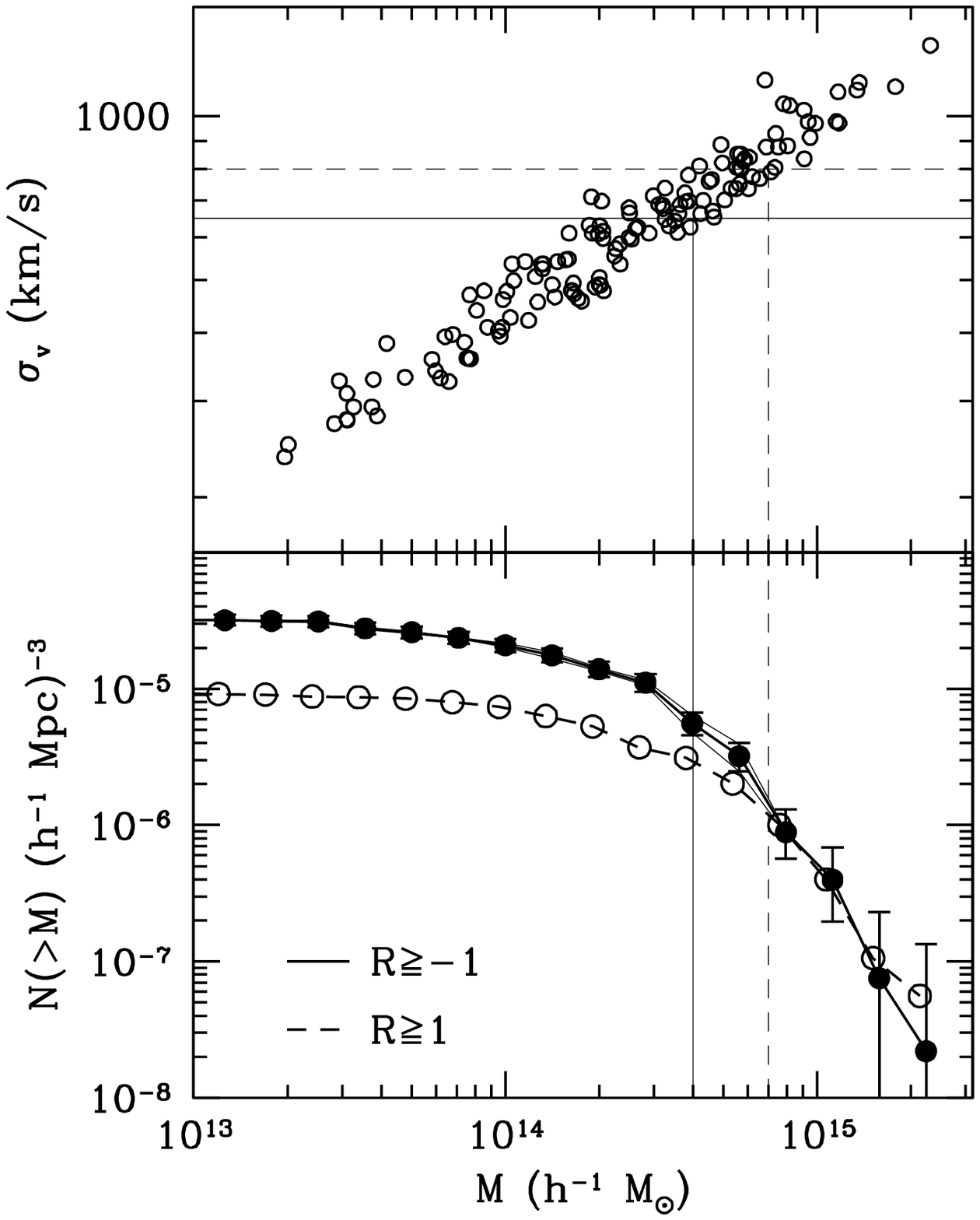}
$\ \ \ \ \ \ $\\
\vspace{9.8truecm}
$\ \ \ $\\
{\small\parindent=3.5mm {Fig.}~2.---
The lower panel shows the cumulative mass
  function for $R\ge -1$ clusters (filled points with solid line) and
  for $R\ge 1$ clusters (open points with dashed line).  Error bars
  represent 1-$\sigma$ uncertainties. The two curves above
  and below the $R\ge -1$ mass function correspond to two different
  ways in which clusters, which appear as multi--peaked in velocity
  space, are treated (see text).  The mass completeness limits for the
  two richness classes (indicated by the vertical lines) are inferred
  from the relation between mass and velocity dispersion, as reported
  in the upper panel and from the corresponding $\sigma_v$
  completeness limits.
}
\vspace{5mm}

We estimate the mass--completeness limit in our sample from that in
$\sigma_v$.  The broad relation between richness and $\sigma_v$ limits
the completeness in velocity dispersion to $\sigma_v\ge 650\vel$ for a
sample with $R\ge -1$ clusters (F96) and to $\sigma_v \ge 800\vel$ for
the sample with $R\ge 1$ clusters (M96). Then, the less scattered
$M$--$\sigma_v$ relation (see the upper panel of Figure~2) is used to
estimate the mass completeness limit. We find $M\magcir 4\times
10^{14}$ \msun for $R\ge -1$ and $M\magcir 7\times 10^{14}$ \msun for
$R\ge 1$.

In the lower panel of Figure~2 we show the resulting cumulative mass
functions, $N(>M)=\int_M^\infty n(M)dM$ for $R\ge -1$ clusters (filled
points) and that for $R\ge 1$ clusters (open points).  As expected,
the curves flatten at small masses due to the effect of the
incompleteness, while they are almost overlapping within the common
completeness range. 
Errors in the mass function have been estimated by taking the
contribution from {\em (a)} the Poissonian uncertainties from the
finite statistics in the G98 sample; {\em (b)} the uncertainty in the
resampling procedure connected with the Poissonian errors in the
richness frequency distribution of the EDCC; {\em (c)} the
uncertainties in the individual cluster mass estimates.
In this plot and in the following, errorbars for
the mass function correspond to the 1-$\sigma$ overall errors
contributed by such three sources.

The ambiguity in how to treat the 17 highly
substructured clusters found in the sample give rise to two different
determinations of the mass function, depending on whether each of them
is treated as a single structure or as disjoined into two objects (cf.
also \S~6 in G98). The resulting mass functions for the $R\ge-1$ case
are plotted in Figure~2 with the lower and upper light continuous
lines, respectively.  It turns out that they are always very close,
the difference being within the statistical uncertainties. In the
following we will refer to the $N(>M)$ plotted with the heavy solid
line, which is computed by assigning equal weights to joined and
disjoined peaks.

\subsection{Comparison with Previous Estimates}

In order to compare our optical virial mass function with those
provided by BC93 and B93, we rescale our cluster mass estimates to the
same 1.5 \h~ radius by using the fitted galaxy distributions (see the
discussion in \S~2.1) and recompute the mass function. The result of
this comparison, which is plotted in Figure~3, shows that our mass
function is intermediate between the two previous determinations.
Following a maximum--likelihood approach, we find the following
expression for our mass
function:
\be
n(M)\,=\,n^*\,\left({M\over M^*}\right)^{-1}e^{-M/M^*}
\label{eq:fit}
\ee 
with $n^*=2.6^{+0.5}_{-0.4}\times 10^{-5}(h^{-1}
Mpc)^{-3}(10^{14}h^{-1}M_\odot)^{-1}$ and $M^*=2.6^{+0.8}_{-0.6} \times
10^{14} h^{-1}M_\odot$.

The mass function by B93 (upper panel of Figure~3) is based on cluster
masses estimated from a homogeneous virial analysis and on the same
resampling procedure that we followed here. Therefore, it is directly
comparable to our results. Since B93 did not provide a normalization
for their mass distribution, we fix it according to the $R\ge 1$
cluster number density provided by M96.  The discrepancy with respect
to B93 can be explained by the fact that the G98 mass estimates are on
average $40\%$ smaller than their estimates. This is explicitly shown
by the dashed curve in the plot, which gives the mass function of B93
after just translating masses by 40\% toward smaller values. As
discussed in detail by G98 (cf. their \S~8), two factors contribute to
this difference. First, the more rigorous algorithm for removing
interlopers gives a smaller $\sigma_v$ by a factor of $10\%$, thus
translating into a $20\%$ difference in mass.  Second, the use of the
surface correction term in the virial theorem (cf. \S~2.1 and Carlberg
et al. 1997b) decreases the mass estimate by about a further $20\%$.

\includegraphics{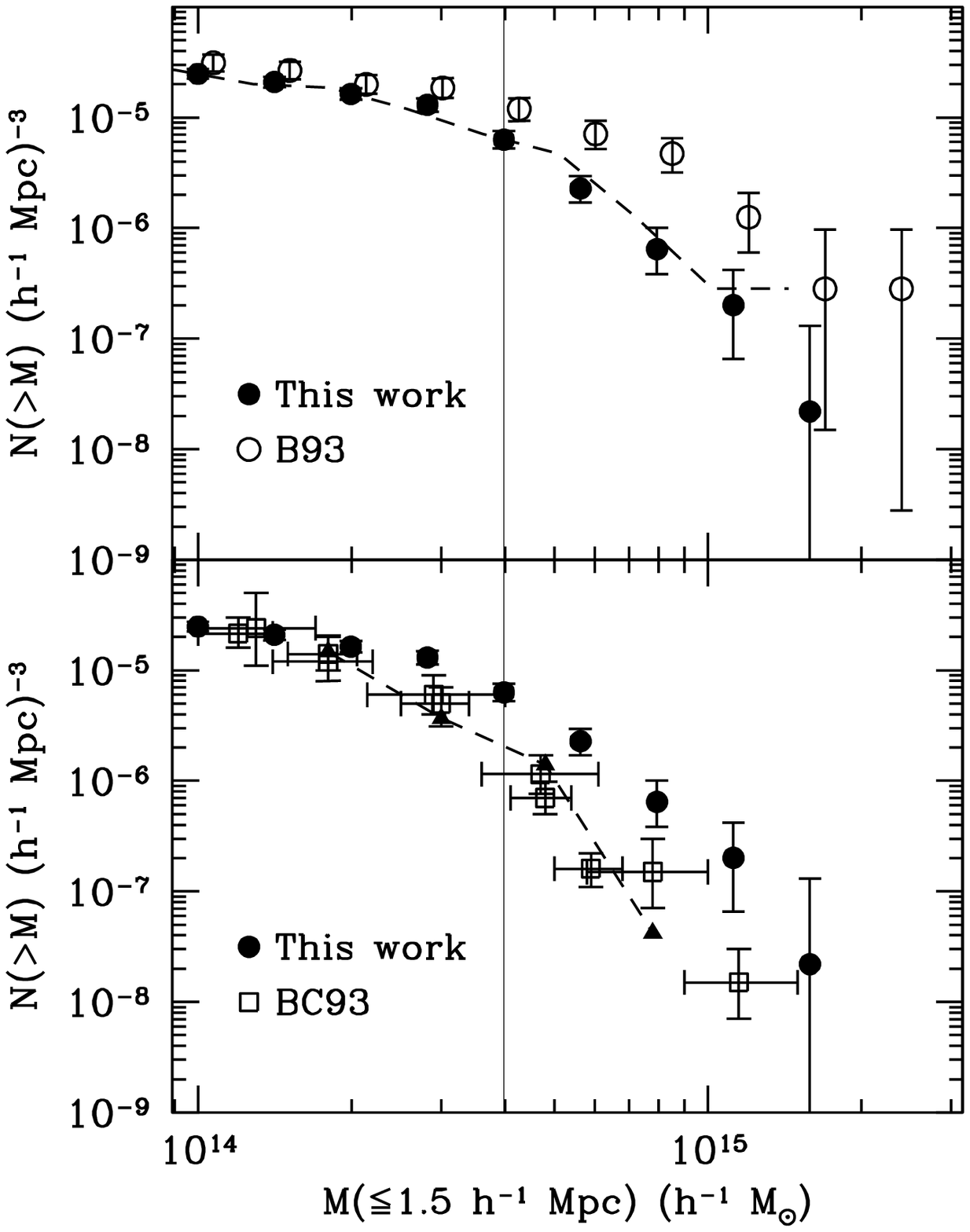}
$\ \ \ \ \ \ $\\
\vspace{9.8truecm}
$\ \ \ $\\
{\small\parindent=3.5mm {Fig.}~3.---
The comparison between this and previous
  determinations of $N(>M)$ by Biviano et al. (1993; upper panel) and
  by Bahcall \& Cen (1993; lower panel; only their determination from
  optical cluster data is reported here).  The vertical line in both
  panels indicates our limiting completeness mass. In the upper panel
  the dashed line indicates the B93 mass function after rescaling
  their masses by 40\% to smaller values. In the lower panel the four
  triangles joined by a dashed line indicate the determination of
  $N(>M)$ we would have obtained from our sample by neglecting the
  scatter in the mass--richness relation (see text).  The plotted
  errorbars for $N(>M)$ represent 1-$\sigma$ uncertainties.
}
\vspace{5mm}

As for the optical data analysis by BC93, they associated the Abell or
EDCC cluster richness to mass according to a suitable one-to-one
scaling relation (cf. their eq.~4; note that this relation falls
within the range, $M(\le 1.5$ \h $)= (0.5-0.7)\,10^{13}N_c$, that we
find to hold for our cluster sample).  
The resulting $N(>M)$ turns out to be somewhat smaller than ours, the
difference being larger at larger masses (cf. the lower
panel of Figure~3). The main reason for this difference is due to the
fact that the procedure by BC93 amounts to neglect any intrinsic
scatter in the mass--richness relation (cf. \S~2.2 about the necessity
of accounting for this scatter).  In order to show the effect of
neglecting this scatter on the BC93 results, we consider the mass
limits corresponding to the $R\ge 0$, $R\ge 1$, $R\ge 2$ and $R\ge 3$
clusters, according to the BC93 mass--richness relation. At these mass
limits, we show the fractions of our resampled clusters (triangles
connected by a dashed line in the lower panel of Figure~3)
corresponding to the four above richness ranges, when neglecting the
contribution from poorer clusters.  This shows the relevance of
properly including the effect of the scatter in the $M$--$R$ relation
and by how much the mass function would have been underestimated
without accounting for it.

In this context, the normalization is only a minor source of
discrepancy between BC93 and our $N(>M)$ since they are comparable at
very small masses ($\sim 1-2 \times 10^{14}h^{-1}M_\odot$) when the
whole cluster sample is considered. However, we stress that, at the
small mass values which are beyond our estimate of completeness limit,
both BC93 and our $N(>M)$ neglect the significant contribution from
the poor galaxy systems which are not recovered in Abell and EDCC
catalogues.

\section{IMPLICATIONS FOR COSMOLOGICAL MODELS}

In order to compare our mass function to predictions of models for
cosmic structure formation, we follow the Press--Schechter (1974, PS)
approach, which provides the following expression for the number
density of local clusters with mass in the range $[M,M+dM]$:
\be
n(M)\,dM\,=\,\sqrt{2\over \pi}\, {\bar \rho \over M^2}\,
{\delta_c\over \sigma_M}\, \left|{d\log \sigma_M\over d\log
M}\right|\, \exp\left(-{\delta_c^2\over 2\sigma_M^2}\right)\,dM\,.
\label{eq:ps}
\ee 
Here $\bar \rho$ is the average matter density and $\delta_c$ is
the linear--theory overdensity for a uniform spherical fluctuation
which is now collapsing; it is $\delta_c=1.686$ for $\Omega_0=1$, with
a weak dependence on $\Omega_0$ for both flat and open geometries
(e.g., Eke et al. 1996). The r.m.s. linear density fluctuation
$\sigma_M$ at the mass scale $M$ is related to the fluctuation power
spectrum according to $\sigma^2_M=(2\pi^2)^{-1}\,\int_0^\infty
dk\,k^2\,P(k)\,W^2(kR)$.  Here $W(x)=3(\sin x-x\cos x)/x^3$ is the
Fourier transform of the window function, assumed to have the top--hat
profile. Accordingly, the mass $M$ of a cluster arising from the
collapse of a fluctuation with typical size $R$ is $M=(4\pi/3)\bar
\rho R^3$.  The PS mass function has been compared with
N--body simulations by several authors (e.g., Eke et al. 1996;
Lacey \& Cole 1996; Borgani et al. 1998; Gross et al. 1998) and has been
generally shown to provide a rather accurate description of the
abundance of virialized halos of cluster size.

Herebelow, we assume for the power spectrum of density fluctuations
the expression $P(k)=A\,k\,T^2(k)$, where the transfer function is
given by
\ba 
T(q)={{\rm ln}(1+2.34 q)\over 2.34 q}\times \ \ \ \ \ \ \ \ \ \ \ \ \
\nonumber\\
\times \left[1+3.89q+(16.1q)^2+(5.46q)^3+(6.71q)^4\right]^{-1/4},  
\label{eq:tk}
\ea
%
%
with $q=k/\Gamma h$.  This expression has been provided by Bardeen et
al. (1986) for CDM models with negligible baryon contribution, once
$\Gamma=\Omega_0 h$ is taken for the shape parameter. In the
following, however, we will take $\Gamma$ to be a free quantity to be
fitted to observations, thus interpreting eq.(\ref{eq:tk}) as a purely
phenomenological expression for the transfer function.  As for the
amplitude of the power--spectrum, it is customary to express it in
terms of $\sigma_8$, the r.m.s. fluctuation amplitude within a
top--hat sphere of 8\h~radius. Therefore, the class of models that we
are considering is represented by three parameters, $\Omega_0$,
$\Gamma$ and $\sigma_8$, either with or without a cosmological
constant term, $\Omega_\Lambda=1-\Omega_0$, to provide flat spatial
geometry. If we further require our model to satisfy the constraint
provided by the four--year {\sl COBE} data (see, e.g., Bunn \& White
1997 and Hu \& White 1997, for flat and open geometry, respectively),
then we would have a one-to-one relation between the shape $\Gamma$
and the amplitude $\sigma_8$ for each value of $\Omega_0$, thus
reducing to two the number of free parameters.

The cluster mass which appears in the PS formula refers to the mass
$M_{vir}$ contained within the virialization radius $R_{vir}$, whose
value depends on the cosmological parameters $\Omega_0$ and
$\Omega_\Lambda$.  Therefore, for the sake of comparison with PS
predictions, the mass function presented in the \S~2.2 should be
suitably re--estimated model by model. If $\Delta_c$ is the ratio
between the mean density within the virial radius and the critical
average density, whose value depends on $\Omega_0$ and
$\Omega_\Lambda$ (e.g., Kitayama \& Suto 1996; $\Delta_c=18\pi^2\simeq
178$ for $\Omega_0=1$), then $R_{vir}^3=3 \Omega_0 M_{vir}/(4\pi \bar
\rho \Delta_c)$. Therefore, the value of $R_{vir}$ depends on
$M_{vir}$, so that we resort to the following recursive procedure to
compute it for each cluster. After taking the mass contained within
the radius $0.002\sigma_v$ as a first guess for $M_{vir}$ (cf.
\S~2.1), we {\em (i)} provide a new estimate of $R_{vir}$ according to
the above expression, and {\em (ii)} estimate the new $M_{vir}$ by
rescaling the previous value to the updated $R_{vir}$, using the
appropriate cluster density profile (cf. \S~2.1).  We explicitly
compute the observational mass function $n(M)$ for
$\Omega_0=0.2,0.4,1$ for both open and flat models, while it is
estimated by linear interpolation for intermediate $\Omega_0$ values.
In our analysis, we applied eq.(\ref{eq:ps}) at $z=0.05$, which
represents the median redshift of the G98 cluster sample.
In any case we find that, for the considered range of models, the mass
function can always be well fitted by the parameteric expression of
eq.(\ref{eq:fit}), with $M^*\simeq 3.5 \times
10^{14} h^{-1}M_\odot$ and normalization varying in a rather small range,
$n^*\simeq (0.9-1.3)\times 10^{-5}(h^{-1}
Mpc)^{-3}(10^{14}h^{-1}M_\odot)^{-1}$.

A further reason of caution when comparing data and model predictions
on $n(M)$ is related to observational uncertainties in estimates of
cluster masses, which are in general not negligible.  G98 provides
mass uncertainties by propagating statistical errors in the estimates
of the velocity dispersion $\sigma_v$ and of the virial radius (cf.
col.~3. of Table~3 in G98).  Such errors are correlated with cluster
masses, being relatively smaller for richer, better sampled clusters.
They range from $\sim 15\%$ at $M\simeq 2\times 10^{15}h^{-1}M_\odot$
to $\sim 35\%$ at $M\simeq 10^{14}h^{-1}M_\odot$.  Therefore, the
measured mass function results from the convolution of the intrinsic
$n(M)$ with mass errors. In order to account for them, we convolve the
PS mass function of eq.(\ref{eq:ps}) with the uncertainties in the
mass estimates, as provided by G98.

We show in Figure~4 the observational $n(M)$, for $\Omega_0=0.4$ and
$\Omega_\Lambda=0.6$, in the mass range where the sample is considered
as complete (cf. \S~2.2).  Also plotted are the predictions for models
with $\Omega_0=0.4$ and $\Omega_\Lambda=0.6$, in order to show the
effect of varying $\sigma_8$ and $\Gamma$ separately (lower and upper
panel, respectively). From the upper panel we show that $n(M)$ is
rather insensitive to the shape of $P(k)$ (see, e.g., White et al.
1993), at least for $M\mincir 10^{15}h^{-1}M_\odot$, where $n(M)$ is
better determined. By varying $\Gamma$ over a rather large range, much
larger than that allowed by the power--spectrum of the observed galaxy
distribution (e.g., Peacock \& Dodds 1994), does not significantly
change the goodness of the fit.  On the contrary, our mass function
represents a strong constraint for the amplitude of the
power--spectrum (cf. the lower panel). The heavy and light curves in
the lower panel show the PS predictions with and without convolving
with mass uncertainties, respectively. Although the effect of such
errors is never large, it turns out that at the smallest mass scale
its size is of the same order of the error bar in the observational
$n(M)$. Therefore, we prefer not to neglect it in our analysis.

\includegraphics{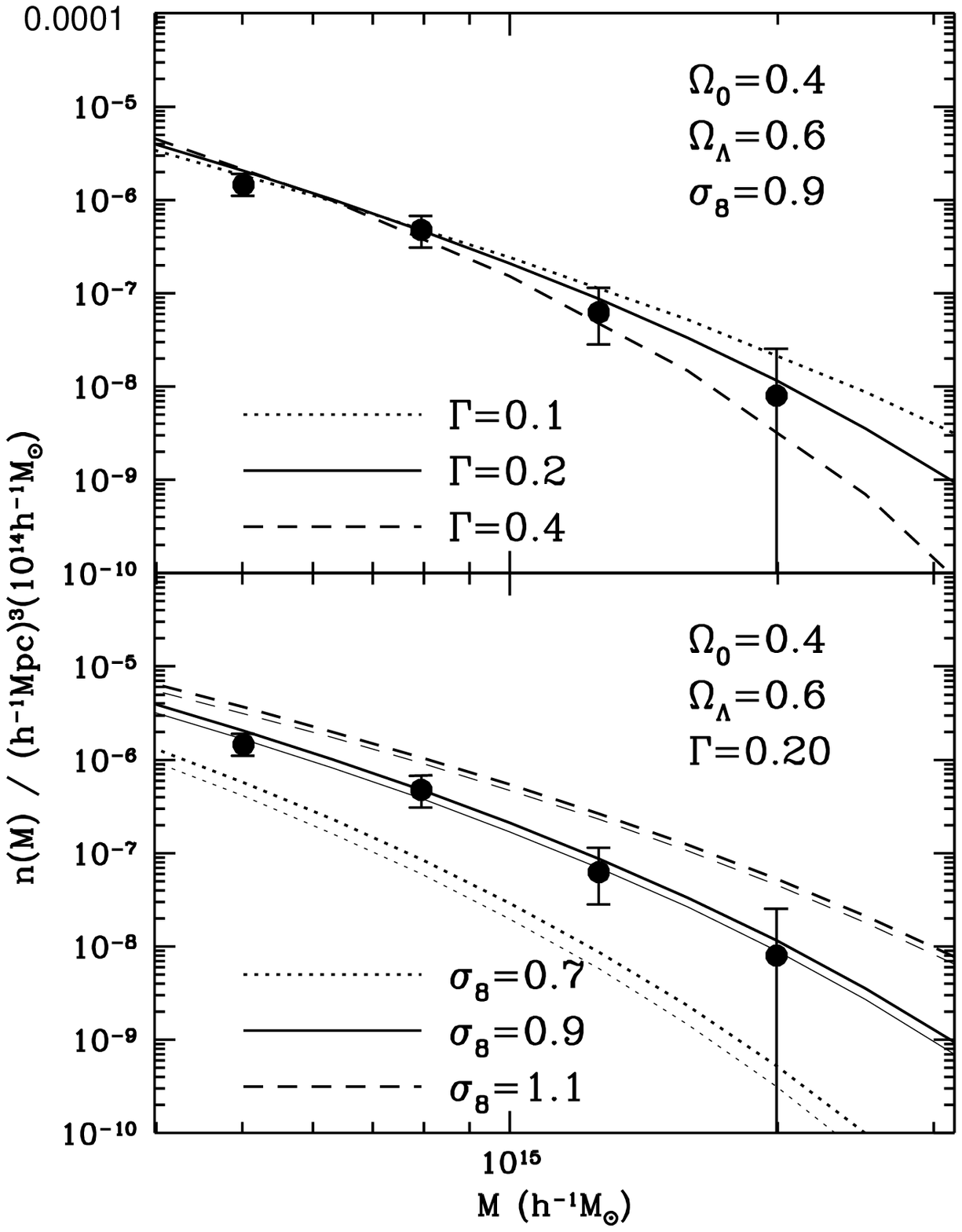}
$\ \ \ \ \ \ $\\
\vspace{9.6truecm}
$\ \ \ $\\
{\small\parindent=3.5mm {Fig.}~4.---
The effect of varying the power--spectrum shape
  parameter $\Gamma$ (upper panel) and $\sigma_8$ (lower panel) on the
  error--convolved Press--Schechter mass function. Heavy and light
  curves in the lower panel are the PS mass function with and without
  convolving the model predictions with the uncertainties in the mass
  estimates. Filled dots represent the observational $n(M)$, with
  masses estimated within the virialization radius for a model with
  $\Omega_0=0.4$ and $\Omega_\Lambda=0.6$.
}
\vspace{5mm}

If we impose the {\sl COBE}--normalization, our observational $n(M)$
provides a relation between $\sigma_8$ and $\Omega_0$ in the model
parameter space. In order to estimate the confidence level for model
rejection, we adopt a $\chi^2$--minimization procedure with respect to
$\sigma_8$ (or, equivalently, with respect to $\Gamma$) for each value
of $\Omega_0$. The probability for a model to be accepted is then
computed as the probability that the observed $n(M)$ comes from the
parent model distribution with Gaussian random variations in
logarithmic units given by the size of the error bars.  As a result,
we find that the 90\% c.l. constraints on the $\sigma_8$--$\Omega_0$
plane can be represented by the fitting expressions
\ba
& & \sigma_8 \,= \, (0.60\pm 0.04)\times
\Omega_0^{-0.46+0.09\Omega_0} ~;~
\Omega_\Lambda=1-\Omega_0 \nonumber \\
& & \sigma_8 \,= \, (0.60\pm 0.04)\times
\Omega_0^{-0.48+0.17\Omega_0} ~;~
\Omega_\Lambda=0
\label{eq:sigom}
\ea
which are accurate within 2\% in the range $0.2\le \Omega_0\le 1$.  
We note that these constraints come from the mass function estimated
above the mass completeness limit which is appropriate for $R\ge -1$
clusters (cf. \S~2.2). If, instead, we use the $R\ge 1$ mass
completeness limit,
$M_{lim}\simeq 7\times 10^{14}h^{-1}M_\odot$, 
then we find $\sigma_8=(0.62\pm 0.08)$ for $\Omega_0=1$,
with similar rescalings as in eq.(\ref{eq:sigom}) for
$\Omega_0<1$. Our results are therefore unaffected by narrowing
the completeness mass range. The only effect is that of increasing the
uncertainties, due to the more limited mass range over which theoretical
predictions are tested.

Our results can be compared with similar results obtained by other
authors from the cluster abundance. White et al. (1993) obtained
$\sigma_8=0.57\pm 0.05$ for $\Omega_0=1$ from the median velocity
dispersion of Abell clusters, as provided by Girardi et al. (1993), and
from the temperature functions by Henry \& Arnaud (1991) and
Edge et al. (1990). Viana \& Liddle (1996) obtained $\sigma_8\simeq
0.6$ for $\Omega_0=1$ by fitting the cluster temperature function by
Henry \& Arnaud (1991) at 7 keV. A somewhat smaller normalization,
$\sigma_8 =0.52\pm 0.04$ for $\Omega_0=1$, has been found by Eke et
al. (1996), also based on the analysis of the cluster temperatures by
Henry \& Arnaud (1991) (see also Markevitch 1998, for a similar result
based on recent ASCA temperature data).  Oukbir, Bartlett \& Blanchard
(1997) found the same result as White et al. (1993), also from the
$X$--ray temperature function. Borgani et al. (1998) obtained
$\sigma_8=0.58\pm0.02$ for $\Omega_0=1$ from the local $X$--ray
luminosity functions provided by Ebeling et al. (1997) and Rosati et
al. (1998). It is remarkable that all such results, although based on
largely different data sets are rather consistent with each other and
converges to indicate that $\sigma_8\simeq 0.6$ for $\Omega_0=1$, with
rescalings for different $\Omega_0$ values which are in general quite
close to those in eq.(\ref{eq:sigom}).

\section{CONCLUSIONS}

In this paper we have presented a new determination of the mass
function $n(M)$ of nearby galaxy clusters.  This analysis is based on
the virial mass estimates by Girardi et al. (1998, G98), for a
sample of 152 Abell-ACO clusters which includes both data from the literature
and the new ENACS data (Katgert et al. 1998).  The extension of this
sample, along with the good agreement found by G98 between optical and
virial masses, makes it the largest data set over which to reliably
estimate the local ($z\simeq 0.05$) cluster mass function.

The main results of our analysis can be summarized as follows. 

\begin{description}

\item[(a)] After applying a suitable resampling procedure to account
  for the lack of volume-- and richness--completeness of the G98
  sample, $n(M)$ is reliably computed for masses larger than
  $M_{lim}\simeq 4\times 10^{14}$\msun. At this mass limit, the value
  of the cumulative mass function is $N(>M)=(6.3\pm 1.2)\times 10^{-6}
  (h^{-1} Mpc)^{-3}$ for masses estimated within the 1.5\h~radius. 
  The corresponding differential mass function is fitted with a
  Schechter--like function [cf. eq.(\ref{eq:fit})], with normalization
  $n^*=2.6^{+0.5}_{-0.4}\times 10^{-5}(h^{-1}
  Mpc)^{-3}(10^{14}h^{-1}M_\odot)^{-1}$ and characteristic mass
  $M^*=2.6^{+0.8}_{-0.6} \times 10^{14} h^{-1}M_\odot$.

\item[(b)] A comparison with previous mass function estimates, also
  based on optical data, by Biviano et al. (1993, B93) and Bahcall \&
  Cen (1993, BC93), shows that our $n(M)$ is intermediate between
  these two, B93 and BC93 providing a larger and smaller estimate,
  respectively (cf. Figure~3).
  
\item[(c)] A comparison with the predictions from models of cosmic
  structure formation demonstrates that our observational mass function
  provides a robust determination of the amplitude of the fluctuation
  power spectrum at the cluster mass scales (cf. Figure~4). After
  suitably convolving the prediction from the Press--Schechter (1974)
  mass function with the observational errors in the mass estimates,
  we determine $\sigma_8$ for {\sl COBE}--normalized models with a
  $\sim 7\%$ uncertainty (at the 90\% c.l.). It turns out that
  $\sigma_8\simeq 0.6$ for $\Omega_0=1$, with appropriate rescalings
  at different $\Omega_0$ values [cf.  eqs.(\ref{eq:sigom})].
 
\end{description}

As a concluding remark, we point out that a precise determination of
$n(M)$ for local clusters, combined with a similar virial analysis of
clusters at higher redshift, $z\simeq 0.3$--0.5, would provide
information about the degree of evolution of the cluster abundance
and, therefore, would allow one to estimate $\sigma_8$ and $\Omega_0$
separately (see, e.g., Bahcall, Fan, \& Cen 1997; Carlberg et
al. 1997a).  The implication of the mass function presented in this
paper on the evolution of the cluster abundance will be discussed in
detail in a future paper.

\acknowledgments S.B. wish to acknowledge SISSA and Osservatorio
Astronomico di Trieste for the hospitality during the preparation of
this paper. This work has been partially supported by
the Italian Ministry of University, Scientific Technological Research
(MURST), by the Italian Space Agency (ASI).

\end{multicols}


\end{document}